\documentstyle{article}

\catcode`\@=11
\def\marginnote#1{}
\newcount\hour
\newcount\minute
\newtoks\amorpm
\hour=\time\divide\hour by60
\minute=\time{\multiply\hour by60 \global\advance\minute
by-\hour}\edef\standardtime{{\ifnum\hour<12
\global\amorpm={am}%
        \else\global\amorpm={pm}\advance\hour by-12 \fi
        \ifnum\hour=0 \hour=12 \fi
        \number\hour:\ifnum\minute<10
0\fi\number\minute\the\amorpm}}
\edef\militarytime{\number\hour:\ifnum\minute<10
0\fi\number\minute}

\def\draftlabel#1{{\@bsphack\if@filesw {\let\thepage\relax
   \xdef\@gtempa{\write\@auxout{\string
      \newlabel{#1}{{\@currentlabel}{\thepage}}}}}\@gtempa
   \if@nobreak \ifvmode\nobreak\fi\fi\fi\@esphack}
        \gdef\@eqnlabel{#1}}
\def\@eqnlabel{}
\def\@vacuum{}
\def\draftmarginnote#1{\marginpar{\raggedright\scriptsize\tt#1}}
\def\draft{\oddsidemargin -.5truein
        \def\@oddfoot{\sl preliminary draft \hfil
        \rm\thepage\hfil\sl\today\quad\militarytime}
        \let\@evenfoot\@oddfoot \overfullrule 3pt
        \let\label=\draftlabel
        \let\marginnote=\draftmarginnote

\def\@eqnnum{(\theequation)\rlap{\kern\marginparsep\tt\@eqnlabel}%
\global\let\@eqnlabel\@vacuum}  }

\def\numberbysection{\@addtoreset{equation}{section}
        \def\theequation{\thesection.\arabic{equation}}}

\def\underline#1{\relax\ifmmode\@@underline#1\else
 $\@@underline{\hbox{#1}}$\relax\fi}

\catcode`@=12
\relax

\numberbysection
\def\beq{\begin{equation}}
\def\eeq{\end{equation}}
\def\bea{\begin{eqnarray}}
\def\eea{\end{eqnarray}}
\def\beqa{\begin{eqnarray}}
\def\eeqa{\end{eqnarray}}

\def\fin{\end{document}}
\begin{document}
\begin{titlepage}
\nopagebreak
\begin{flushright}
LPTENS-99/41\\
hep--th/9911008
\\
August   1999
\end{flushright}
\vglue 2.5 true cm
\begin{center}
{\bf BACKLUND TRANSFORMATIONS IN  10D SUSY    
YANG--MILLS THEORIES} \\
{\it (with a  preamble on the birth of  supersymmetry)}\\
\bigskip
{\sl Contribution to Y. Golfand Memorial Volume\\
``Many faces of Superworld'', World Scientific. } \\
\bigskip 
{ Jean--Loup GERVAIS,\\
 \bigskip {\footnotesize Laboratoire de
Physique Th\'eorique de l'\'Ecole Normale Sup\'erieure
\footnote{UMR 8549:  Unit\'e Mixte du Centre National de la Recherche Scientifique, et de 
l'\'Ecole Normale Sup\'erieure. },\\ 24 rue Lhomond, 75231 Paris C\'EDEX
05, ~France.}}
\end{center}
\bigskip 
\baselineskip .4 true cm
\noindent
\begin{abstract}
A B\"acklund transformation is derived for the Yang's type (super) equations  
 previously derived (hep-th/9811108) by M. Saveliev and the author, from 
 the ten dimensional super Yang-Mills field equations  in an on--shell light cone
 gauge.  It is shown to be based upon a particular gauge transformation satisfying  nonlinear
conditions which ensure that  the equations retain the same form.
These Yang's type field equations are shown to be precisely such that they automatically
provide a solution of these conditions. This  B\"acklund transformation is similar to the one
proposed by A. Lesnov for self-dual Yang-Mills in four dimensions. In the introduction a personal
recollection  on the birth of supersymmetry is given.   
\end{abstract}
\vfill
\end{titlepage}

\section{ Remembering the early times. }
{ \it A general history  of the birth of  supersymmetry may be found elsewhere in this
volume  (see ``{\sl Revealing The Path To Superworld}'' by M.S. Marinov), and I will not try to
give a complete overview. 
  As is well known,  supersymmetry started long before  
the Iron Curtain was dismanteled, and thus  came into existence separately in the West and 
the former Soviet Union.  This volume is (rightly) dedicated to the pionnering contribution of Yuri
Abramovich Golfand. With  Evgeny Likhtman, he took many basic initial steps towards supersymmetry
as we know now.  An account of this story may be found in Shifman's Introduction to this book, and
in  Likhtman's ``{\sl Notes of a Former Graduate Student}'' elsewhere in this volume. 
Regretfully, I never had the chance to meet Golfand in person. I hope very much that this volume
will play an important role in the proper recognition of  his memory.

 In order to speak about what I know personally, I will recall that, in  the West, superalgebras
were first considered by A. Neveu, J. Schwarz and P. Ramond   as a basic tool to eliminate ghosts
from the spinning string theories, when they first introduced them ---they were  thus initially
refered to as supergauges. These authors used the covariant harmonic oscillator approach, the only
known technics at that time, without  field theoretic interpretation.    
Preparing the present
text  brought back wonderful memories of  the time when my long lasting collaboration and
friendship with Bunji Sakita \footnote{then a Visiting Professor both at the Institut des Hautes Etudes, in  Bures-sur
Yvette,  and at the Laboratoire de Physique Th\'eorique et Hautes Energies,  University of Orsay,
France.} began. In 1970-71 we   started to develop  the  world--sheet interpretation  of the
spinning string  ---unknown at that early time--- extending earlier discussions of the purely
bosonic case begun by H. Hsue B. Sakita and M. Virasoro. We recognized that the Neveu-Schwarz
Ramond models include world--sheet two--dimensional Dirac spinor fields in addition to the world
sheet scalar fields common with the Virasoro model.  We showed that the supergauges of the NS-R
models correspond to the fact that  the  two dimensional world--sheet Lagrangian is invariant under 
transformations with anticommuting parameters which mix the scalar and spinor fields. In the
West, this gave the first example of a supersymmetric local Lagrangian. 

This supersymmetry  was 
closing only on--shell for lack of auxiliary fields, and was two dimensional. In the West, the
  problems of proceeding off shell and to four dimensions were  initially solved by Wess and Zumino,
as is well known.  

These are the historical facts simply  stated.   It is worth trying to give a more personal picture
of the birth  of supersymmetry. For me, the story really begins during the year 1968-69 which was a
sort of  turning point. I was just returning from two years of postdoc at New York University, where
I had met B. Zumino (then the head of the Theory Group), K. Symanzik, W. Zimmermann  (at that time
permanent faculty members) and J. Wess (a visitor for one year). Before that my interest was
mostly on dispersion relations, Regge poles, and S-matrix theory, but at NYU, I had been  fully
converted to local field theory, and much impressed by the power of symmetries in that context, be
they local or global. Of course, this year saw the beginning of string theory which was, however
initially developed using the covariant operator method within the context of S-matrix theory,
giving what looked like a realisation of G. Chew's program. On the other hand, local field theory
also made wonderful progress on its own.  The main problems of that time  were
the Adler-Bell-Jackiw anomaly, the spontaneous breaking of symmetries and the quantization of
Yang-Mills theory. For the latter, the work of L. Faddeev and V. Popov was gradually becoming more
and more popular. It is hardly necessary to say that these topics now belong to textbooks.  At that
time the French Government was  very generous with temporary positions, and a handful of key
visitors  came for long visits
 \footnote{at the Laboratoire de Physique Th\'eorique et Hautes Energies of  Orsay
(France) where I was working permanently then,} during that wonderful year: D.Amati, the late
Benjamin Lee,
 T. Veltman, and B. Zumino. I drew much inspiration from
the very stimulating atmosphere they created, together with the more senior permanent members. In
particular, with Amati, and Bouchiat I devised the now standard method to compute loops in string
theories, using coherent states, and with B. Lee, I showed how to correctly quantize the linear
$\sigma$ model in the phase where the spontaneous symmetry breaking takes place \footnote{ It seems
that G. t'Hooft ---then a student a Cargese during the following summer--- drew much inspiration for
quantizing massive Yang-Mills theory,  from B. Lee's lecture on spontaneous symmetry breaking.}. 

This is all to say that, when I first met Bunji Sakita in  fall 1970, I was fully
motivated to apply field theory technics to string theories.   Moreover, the year before I 
had shown that the integrand of the Veneziano model is equal
to the vacuum expectation of a product of scalar field which are functions of the Koba Nielsen
variables. This result, similar to an independent   and better known work  by S. Fubini and
G. Veneziano,  was indeed 
 a strong hint of the world-sheet field theory aspect of string theory. 
This viewpoint is now a common
place, but at that time it was not at all popular among string theorists.
A large majority prefered the  operator method, which  achieved  striking technical
success. 
 
Before we met, Sakita and his collaborators had already made important progress in developing
world-sheet field theory technics using path integrals.   On the one hand, H. Hsue, B. Sakita, and
M. Virasoro had shown how the analog model of H.B. Nielsen could be derived from the path integral
over a free scalar 2D field. On the other hand, Sakita came with the draft of an article where he
had started to discuss Feynman-like rules for the Veneziano model using the  
factorisation of path integrals over sliced Riemann surfaces.   There were many basic problems
left, and at the beginning, we spent a lot of time establishing a general scheme.   
This complicated work was not so well-received, although it contains many precursive results. For
the following, the most important point was that we made an essential use of the conformal
invariance of the path integral representation over  scalar free fields in two dimensions.  
Although we did not really consider the gauge-fixing problem at that time,  we were pretty much
convinced that  conformal invariance of the  free world-sheet action is at the origin of the
negative normed state elimination.  On spring  when we came accross the first article of A. Neveu
and J. Schwarz, this motivated us to systematically discuss conformal field theories in two
dimensions, as a way to classify string theories, by defining what we called irreducible fields
---now known as primary fields, following A. Belavin, A. Polyakov, and A.B. Zamolodchikov. 
Considering only quadratic actions we recognized that only spin-zero and
spin-one-half  fields were possible, covering all existing critical  string models of today. This
was all very well except for one fact: the Neveu-Schwarz model has more ghosts and needs an
additional negative-normed-state killing mechanism as compared with the Veneziano model. This had
motivated these authors, as well a Ramond in his seminal work, to introduce in the operator
formalism a set of operators whose anticommutators gave the Virasoro generators. The visit of
Sakita in France was about to terminate, and he became  busy with moving with his  family, but I
quickly started to look for the possible symmetry of the action that would be the origin of this
additional ghost killing. From the form of the NS-R generators it was immediately clear that the
transformed of the boson had to be a fermion and vice versa.  It was not difficult to envisage that
the action could be invariant, except for the mixing between commuting and anticommuting fields
which made everything very confusing. After many hesitations, Sakita and I solved the problem by
introducing symmetry transformations with anticommuting parameters, from which the supersymmetry of
the world-sheet NS-R action followed  very simply. The paper was completed and thus for us
supersymmetry was born (August 1971) just the day before Sakita departed France to fill his new
prestigious position at City College New-York.      

 At that time, the use of anticommuting c-numbers was not well
appreciated in our communities, and this work did not get much attention in general. In December
1971 Sakita delivered a talk about it at the {\sl Conference on Functional Methods in Field Theory
and Statistics}  at the Lebedev Institute in Moscow, organised by E. Fradkin.  On the way back,
Sakita stopped over in Paris, and we wrote a summary of our ideas for the proceedings which was
sent to the organisers and  circulated as a preprint ---the complete proceedings them--selves were
never published \footnote{We later published our text  in  {\sl Quantum Field Theory and Quantum
Statistics}, Essays in honour of the sixtieth birthday of E.S. Fradkin, vol 2, p. 435, Adam Hilger,
1987.}. Through this, there  was some early communication of our work to the soviet  scientific
community.  

Scientifically we separated at a very unfortunate moment of our research program. At that time of
course there was no email. Phone was expensive and airmail slow. Moreover Sakita became busy with
his new life and responsabilities. I did not push very hard further in the direction of
Supersymmetry to my regret. Other problems seem more pressing. In the mean time the
Nambu-Gotto action and the Goddard Goldstone Rebbi and Thorn light-cone quantization had come
out.  Sakita and I showed \footnote{This work was initiated during my one-month visit at City
College during October 1972.} how the latter may be recovered using path integral with the former
action using   the   Faddeev-Popov method in order to handle reparametrisation invariance.  This
work raised much  interest and criticism
\footnote{in particular from the referee and at a seminar which Sakita gave at the Institute in 
Princeton on December 1972, }. The main objection was that our gauge depends upon the external
sources, and thus is not easily factorizable, in contrast with our previous path integral
formulation. We tried hard to understand what was going on but failed. The answer was given by
S. Mandelstam: the light-cone gauge is not conformally invariant, so that there is only one 
(prefered) parametrisation where factorization holds with our gauge fixing. With this
parametrisation  one sees strings (with lengths equal to their respective 
$p_+$) which split and join, and our work played a key role     
  for Mandelstam's subsequent discussion of scattering amplitudes, which led to light-cone string
field theory. 

On the other hand, there
remained the problem of obtaining our free world-sheet action of the NS-R models by gauge fixing
from a local Lagrangian. This problem, first considered  by  Iwasaki and K. Kikkawa,  
played a key role in the subsequent development of supersymmetry  
\footnote{ At that time there were both at
City College and had much interaction with Sakita. }.   In spring 1973, Sakita visited
the Niels Bohr Institute and I made a trip there to meet with him.  On the way back home, he went to
CERN and gave a talk where Zumino was present. Sakita reviewed the work of Iwazaki and  Kikkawa.
Later on it appeared that this seminar and a later conversation with Zumino played a key role in
leading Wess and Zumino to begin their seminal work on supersymmetry. I also remember that the
latter author asked me questions about our works on various occasions. After that the whole subject
suddenly exploded, and our contribution was temporarily  forgotten for lack of
reference \footnote{Our original paper was nevertheless reprinted in the first volume of {\sl
SUPERSTRINGS The first 15 years of superstring theory} edited by J. Schwarz.}, given the fact that
we had turned to other research directions.     

Many other basic developments were initiated in those wonderful times. In particular, and since this
article deals with the  super Yang-Mills theories in ten dimensions which appear in the zero slope
limit of type I string theory, it may be worth recalling that I was at the same time   
collaborating with A. Neveu on the zero-slope limit of string theories
\footnote{ which had just been developed for the purely bosonic case by J. Scherk, following a
remark by R. Omnes at a lunch at Orsay where I was also present, and quickly extended by A. Neveu
and J. Scherk to spinning strings.  Our work was dug out  two decades later when Bern
and Kossover showed that string inspired rearrangments of Feynman graphs give tremendous
simplifications in high order perturbation of gauge theories.}.} 
 
\section{Introducing the present work. }
Three decades later, it obvious  that supersymmetry has played a key role in theoretical physics,  
although it is  so badly broken at accessible energies that it is not yet verified
experimentally. Generally speaking, supersymmetric theories seem to enjoy striking properties
which led to remarkable developments. In particular  non perturbative
results  have been derived in theories with  extended supersymmetries which show a striking 
connection between exact integration and supersymmetry.   The present work initiated in
collaboration with the late Misha Saveliev\cite{GS98} goes along this line, being connected with the
exact integration of classical super Yang-Mills in ten dimensions. Here as in many recent
theoretical advances, supersymmetry plays a key role. Thus it is proper to present this work as a
tribute to the  Yuri Golfand Memorial Volume. 

 Shifman's introduction to the present volume bears a striking testimony of the hardships which
Yuri Golfand underwent  in the terrible pressure of the communist system. This is
fortunately gone by now. It is  a very nice that it is 
possible to collabarate freely with the russian scientists who chose to remain in their native
country, although the practical life there has become so difficult that  some of them have died
prematurely.  

At this point  another historical note may be in order. My work with Sakita  on functional approach
to string theories did not deal with quantum anomalies. The basic reason is   that there	 is no way
to regularize the Nambu-Gotto action while preserving reparametrization invariance.  This problem 
 was nicely cured   a decade later by  S. Polyakov using the action of L. Brink P. Di Vecchia and
P. Howe. This led me, with A. Neveu and others to  extensive studies of  the exact quantum
solution of the Liouville theory. This theory is completely integrable classically and is a
particular case of the  Toda theories which were nicely solved by A. Lesnov and M. Saveliev.
With A. Bilal, I showed that they obey \footnote{speaking only about the conformal ones,}  W
symmetries in general. This raised my interest in Toda theories and in integrable theories in
general. This is how I was led me to collaborate with M. Saveliev untill his unfortunate death. It
is now  apparent that theories with enough local supersymmetries are the higher dimensional
analogues of two dimensional conformal/integrable theories. In particular it has been
known\cite{W86,AFJ88}  already for some time  that ten dimensional supersymmetric Yang--Mills are
similar to self-dual Yang-Mills theories in four dimensions in the sense that the field equations
are equivalent to flatness conditions. More recently, the interest was revived into  (suitably
reduced) ten dimensional supersymmetric Yang--Mills theories in the large
$N$ limit since they have been actively  considered in  the search for the M theory (see e.g
refs\cite{BFSS96,DVV97}). This has motivated us to return to the use of  flatness conditions
in superspace in order to derive non trivial classical solutions.   Our initial
idea was to try to apply method inspired by the ones of Lesnov and Saveliev to them. However,
different approaches, more closely inspired from self-dual Yang-Mills in four dimensions have
turned out to be more fruitful \cite{GS98} \cite{G99-1} \cite{G99-2}.   
  
Let us first recall some standard formulae in order to establish the notations. 
In ten dimensions the dynamics is specified by the standard action 
\beq
S=\int d^{10} x {\> \rm Tr   }
\left\{
{-1\over 4}Y_{mn}Y^{mn}
+{1\over 2}\bar \phi\left(\Gamma^m \partial_m \phi+\left[X_m,\, \phi\right]_- \right)\right\}, 
\label{action}
\eeq
\beq
Y_{mn}=\partial_mX_n-\partial_nX_m +\left[X_m,\, X_n\right]_-.
\label{F0def} 
\eeq
The notation is as follows. $X_m(\underline x)$ is the vector potential, $\phi(\underline x) $ is
the Majorana-Weyl spinor. Both are matrices in the adjoint representation of the gauge group 
${\bf G}$.  Latin indices
$m=0,\ldots 9$ describe Minkowski components.  Greek indices $\alpha=1,\ldots 16$ denote spinor
components. We will use the superspace formulation with odd coordinates $\theta^\alpha$. The  super
vector potentials, which are valued in the gauge group, are noted  
$A_m\left(\underline x,\underline \theta\right)$, $A_\alpha\left(\underline x,\underline
\theta\right)$. It has been shown \cite{AFJ88} that we may
remove all the additional fields and uniquely reconstruct the physical fields $X_m$, $\phi$ from
$A_m$ and $A_\alpha$ if we impose the condition $\theta^\alpha A_\alpha=0$ on the latter.

With this condition   the field equations derived
from the Lagrangian  \ref{action} are equivalent \cite{W86} \cite{AFJ88} to the flatness conditions 
\beq
F_{\alpha \beta=0}, 
\label{flat}
\eeq
where $F$ is the supercurvature 
\beq
F_{\alpha \beta}=D_\alpha A_\beta+D_\beta A_\alpha+\left[A_\alpha,\, A_\beta\right]_++
2\left(\sigma^m\right)_{\alpha\beta}A_m.  
\label{curdef}
\eeq
 $D_\alpha$ denote the superderivatives
\beq
D_\alpha=\partial_\alpha-\left(\sigma^m\right)_{\alpha \beta} 
\theta^\beta {\partial_m}, 
\label{sddef}
\eeq
and we use the Dirac matrices 
\beq
\Gamma^m=\left(\begin{array}{cc}
0_{16\times16}&\left(\left(\sigma^m\right)^{\alpha\beta}\right)\\
\left(\left(\sigma^m\right)_{\alpha\beta}\right)&0_{16\times16}
\end{array}\right),\quad  
\Gamma^{11}= \left(\begin{array}{cc}
1_{16\times16}&0\\0&-1_{16\times16}\end{array}\right).
\label{real1}
\eeq
The physical fields  appearing in equation \ref{action} are reconstructed from the superfields $A_m$
$A_\alpha$  as follows. Using the Bianchi identity on the super curvature one shows that 
one may write  
$$
F_{\alpha m}=\left(\sigma_m\right)_{\alpha \beta}\chi^{\beta}. 
$$ 
Then  $X_m$, $\phi^\alpha$ are, respectively,  the zeroth order contributions in the expansions of 
$A_m$ and $\chi^{\alpha}$ in powers of the odd coordinates $\theta$.

Throughout the paper, it will be convenient to use the following particular realisation: 
\beq
\left(\left(\sigma^{9}\right)^{\alpha\beta}\right)=
\left(\left(\sigma^{9}\right)_{\alpha\beta}\right)=
\left(\begin{array}{cc}
-1_{8\times 8}&0_{8\times 8}\\
0_{8\times 8}&1_{8\times 8}
\end{array}\right)
\label{real2}
\eeq
\beq
\left(\left(\sigma^{0}\right)^{\alpha\beta}\right)=-
\left(\left(\sigma^{0}\right)_{\alpha\beta}\right)=
\left(\begin{array}{cc}
1_{8\times 8}&0_{8\times 8}\\
0_{8\times 8}&1_{8\times 8}
\end{array}\right)
\label{real3}
\eeq
\beq
\left(\left(\sigma^{i}\right)^{\alpha\beta}\right)=
\left(\left(\sigma^{i}\right)_{\alpha\beta}\right)=\left(\begin{array}{cc}
0&\gamma^i_{\mu,\overline \nu}\\
\left(\gamma^{i\, T}\right)_{\nu,\overline \mu}&0
\end{array}\right),\quad  i=1,\ldots 8. 
\label{real4}
\eeq
The convention for greek letters is as follows: Letters from the beginning of the alphabet run from
1 to 16. Letters from the middle of alphabet run from 1 to 8. In this way,  we shall separate
the two spinor representations of $O(8)$ by rewriting $\alpha_1,\ldots, \alpha_{16} $  as 
$\mu_1,\ldots, \mu_8, \overline \nu_1,\ldots, \overline \nu_8$. 
   
Using the above explicit realisations on sees that the equations to solve take the form  
\begin{eqnarray}
D_\mu A_\nu+D_\nu A_\mu +\left[A_\mu,\,
A_\nu\right]_+&=&2\delta_{\mu\nu}\left(A_0+A_9\right),\label{dynuu}\\
D_{\overline \mu} A_{\overline \nu}+D_{\overline \nu} A_{\overline \mu} +\left[A_{\overline \mu},\,
A_{\overline \nu}\right]_+&=&
2\delta_{{\overline \mu}{\overline \nu}}\left(A_0-A_9\right), \label{dyndd}\\  
D_{ \mu} A_{\overline \nu}+D_{\overline \nu} A_{ \mu} +\left[A_{\mu},\,
A_{\overline \nu}\right]_+&=&-2\sum_{i=1}^8 A_i\gamma^i_{\mu,\overline \nu}.\label{dynud}
\end{eqnarray}
At this point one makes use of the fact that equations \ref{dynuu} and \ref{dyndd} are,
respectively the integrability conditions of the equations   
\beq
\left(D_\mu -A_\mu\right)R_+=0 ,\quad  \left(\partial_+-A_+\right) R_+=0, 
\label{R+def}
\eeq
\beq
\left(D_{\overline \mu } -A_{ \overline \mu}\right)R_-=0,\quad 
\left(\partial_--A_-\right) R_-=0, 
\label{R-def}
\eeq
where $R_\pm$ are superfields valued in the gauge group. 
We let from now on 
\beq
A_{\pm}={A_0}\pm { A_9},\quad 
\partial_{\pm}={\partial\over \partial x^0}\pm {\partial\over \partial x^9}. 
\label{pmdef}
\eeq
 A straightforward computation shows that equations \ref{dynud} become 
$$
D_{\overline\nu}\left( {\bf R}^{-1}\, D_{\mu}\, {\bf R}\right) =-2\sum_{i=1}^8\tilde{A}_i
\gamma^i_{\mu,\overline\nu}
$$
$$
{\bf R}\equiv R_+R_-^{-1},\quad 
\tilde{A}_i\equiv R_-(A_i+\partial_i) R_-^{-1}. 
$$ 
In practice, given ${\bf R}$, we may derive  the field $\tilde A_i$, if the following conditions
hold
\beq
\sum_{\mu \overline \nu  }D_{\overline\nu}\left( {\bf R}^{-1}\, D_{\mu}\, {\bf R}\right)
\gamma^{ijk}_{{\mu \overline \nu  }}=0,
\quad 
1\leq  i<j< k\leq 8.
\label{Dcond}
\eeq 
These are complicated  non linear $\sigma$ model type equations in superspace which so far could
not be handled. This is basically why these reasonings did 
not allow yet  to construct any explicit nontrivial physically meaningfull solution. Conditions 
\ref{Dcond} only provide  a procedure \cite{AFJ88} for obtaining infinite series of nonlocal, and
rather complicated conservation laws. 
\section{Yang's form of the field equations}
In this section, we repeat a previous discussion\cite{GS98} for completeness.   
\subsection{A usefull on-shell gauge}
Under gauge transformations, we have  
$$
R_\pm \to R_\pm \Lambda, \quad 
A_m\to \Lambda^{-1}\left(A_m+\partial_m\right) \Lambda,\quad 
A_\alpha\to \Lambda^{-1}\left(A_\alpha+D_\alpha\right) \Lambda. 
$$
Thus ${\bf R}$ is gauge invariant. If $\Lambda=R_-^{-1}$, we get 
\beq
A_+\to{\bf R}^{-1}\partial_+{\bf R} ,\quad  A_\mu \to 
{\bf R}^{-1}D_\mu{\bf R},\quad 
 A_i\to \widetilde A_i, 
\label{osgauge}
\eeq   
$$
A_-\to 0, \quad A_{\overline \mu }\to 0. 
$$
Thus, if the field equations  are satisfied there exists a gauge (on shell) such that  
$A_-=A_{\overline \mu }=0$. After this gauge choice, the flatness conditions
\ref{dynuu}--\ref{dynud} boil down, respectively,  to
\begin{eqnarray}
D_\mu A_\nu+D_\nu A_\mu +\left[A_\mu,\,
A_\nu\right]_+&=&4\delta_{\mu\nu}A_0,\label{dynuu0}\\
0&=&0,\label{dyndd0}\\
D_{\overline \nu} A_{ \mu} &=&-2\sum_1^8 \widetilde A_i\gamma^i_{\mu,\overline \nu}. 
\label{dynud0}
\end{eqnarray}
The last  mixed ones which in general lead to the complicated conditions \ref{Dcond} have become
linear, and  their general solution may be derived in closed form\cite{GS98}.  
\subsection{Dynamical equations for the superpotential field  $\Phi$.}
It follows from the Dirac algebra that the $\gamma$ matrices satisfy the equations 
\beq
\gamma^i\gamma^{j T}+\gamma^j\gamma^{i T}=2\delta_{ij},\quad 
i,\,j=1,\ldots, 8.
\label{o8gam}
\eeq 
We may choose $\gamma^8=1$. Then it follows that the other matrices, i.e. $\gamma^i$,
$i=1,\ldots,7$ are antisymmetric. With this choice of realisation, it is convenient to separate
the symmetric and antisymmetric combinations of   equations
\ref{dynud0}. This gives 
\beq
D_{\overline \nu} A_{ \mu}+
D_{\overline \mu} A_{ \nu}   =-4\widetilde A_8\delta_{\mu,\overline \nu}
\label{symeq}
\eeq
\beq
D_{\overline \nu} A_{ \mu}-
D_{\overline \mu} A_{ \nu}   =-4\sum_1^7 \widetilde A_i\gamma^i_{\mu,\overline \nu}
\label{asymeq}
\eeq
By convention    greek letters with and without overline take the same numerical values, so
that, for instance, $\gamma^8_{\mu \overline \mu}=1$.  Next, with the
present particular realisation, one may verify that $\left[D_{\overline \mu},\,D_{\overline \nu}
\right]_+=2\delta_{\overline \mu, \overline \nu}\partial_+$. Thus  it follows  from equations
\ref{symeq} that there exists a superfield
$\Phi$, such that
\beq
A_\mu=D_{\overline \mu} \Phi. 
\label{Phidef}
\eeq
Then one finds that   
\beq
 \tilde A_8=-2\partial_+\Phi.
\label{a8res}
\eeq 
Finally, one may eliminate $A_{\overline \mu}$ from all the remaining dynamical equations 
using equation \ref{Phidef}. This gives  
\beq
\left[D_{\overline \nu},\, D_{\overline \mu}\right]_- \Phi
=-4\sum_1^7 \widetilde A_i\gamma^i_{\mu,\overline \nu}
\label{self}
\eeq
\beq
D_\mu D_{\overline \nu}\Phi+D_\nu D_{\overline \mu}\Phi + 
\left[D_{\overline \mu}\Phi,\,  D_{\overline
\nu}\Phi\right]_+=4\delta_{\mu \nu} A_0. 
\label{nonleq}
\eeq
The general solution of equations \ref{self} has been given in closed form\cite{GS98}. 
Equations \ref{nonleq} are similar to Yang's equations. A partial class of solutions of these
equations may be derived\cite{GS98} using methods similar to the ones developed\cite{LS89} for
self-dual Yang-Mills theories in four dimensions. So far, however, it has not been possible to
derive a solution of equation \ref{nonleq} which also satisfies equations \ref{self}. Thus we have
still failed to solve the full Yang-Mills equations. In the search for more general solutions of
\ref{nonleq}, it is clear that the existence of B\"acklund transformations for the solutions of
these equations  may be very useful. In the present discussion we  concentrate upon this topic,
leaving aside the much harder problem of deriving simultaneous solutions of equations \ref{nonleq}
and \ref{self}. We will have more to say at the end. 

Before beginning the discussion, it is useful to present a generalisation of an argument given
earlier\cite{GS98} in the particular case where there is no dependence upon $x_1,\ldots, x_8$, 
since it turns out to be crucial for the existence of the B\"acklund transformations we have in
mind.  The point is to show that equations \ref{nonleq} are equivalent to a set of differential
conditions which are first order in the superderivatives. Using the fact that 
$$
\left[D_{\mu },\, D_{\overline \nu}\right]_+=-2\gamma^i_{\mu, \overline \nu}\partial_i
$$
we rewrite equation \ref{nonleq} as  
\beq
D_{\overline \nu} D_\mu \Phi+ D_{\overline \mu}D_\nu\Phi
+ g \left[D_{ \overline \mu}\Phi,\, D_{ \overline \nu}\Phi\right]_+=-4\delta_{\mu \nu}\left(
A_0+\partial_8 \Phi\right)
\label{nonleq2}
\eeq
Next let us consider the superfield $\Omega_{\overline \mu}$ defined by    
$$
\Omega_{\overline \mu}
 =D_\mu\Phi-{1\over 2 }\left[\Phi, D_{\overline \mu}\Phi\right]_-
$$
Using equations \ref{nonleq2} one finds that  
$$
D_{\overline \nu}\Omega_{\overline \mu}+
D_{\overline \mu}\Omega_{\overline \nu}
=-2\delta_{\mu \nu}\left(
2 A_0+2\partial_8 \Phi+{1\over 2} \left[\Phi,\,
\partial_-\Phi\right]_- \right)
$$
It thus follows that  we may let $\Omega_{\overline \mu}=D_{\overline \mu}\Omega$. 
Finally equations \ref{nonleq2} are equivalent to the conditions 
\beq
D_\mu\Phi=D_{\overline \mu}\Omega+{1\over 2}\left[\Phi,\, D_{\overline \mu}\Phi\right]_-
\label{omeq}
\eeq
with 
 \beq
A_0=-\partial_8 \Phi -{1\over 2}\partial_-\Omega
-{1\over 2 }\left[\Phi,\, \partial_-\Phi\right]_-
\label{Aodef}
\eeq  

At this point it is interesting to recall the  four dimensional Yang equations which arose in
solving self--dual (purely bosonic) Yang--Mills in four dimensions. For this, we closely
follow an earlier  review \cite{LS89}.  There are two bosonic complex coordinates $z$, $y$ and their
conjugate
$\bar z$, $\bar y$. One may start from the equations  (Indices mean derivatives)
$$
\left(G_zG^{-1}\right)_{\bar z}+\left(G_yG^{-1}\right)_{\bar y}=0,
$$
where $G$  is in the adjoint representation of the gauge group. This is partially solved by letting 
\beq
G_{ z}G^{-1}=f_{\bar y},\quad G_{ y}G^{-1}=-f_{\bar z}.
\label{fdef}
\eeq
which leads to the consistency condition 
\beq
f_{\bar z z}+f_{\bar y y}+\left[f_{\bar y},\, f_{\bar z}\right]_-=0.
\label{feq}
\eeq
In order to draw a parallel with our case, let us recall that, according to equations 
\ref{osgauge}, \ref{Phidef}, we have 
\beq
A_\mu={\bf R}^{-1}D_\mu {\bf R}=D_{\overline \mu} \Phi. 
\label{R-Phi}
\eeq
There is a similarity between equations \ref{nonleq} and \ref{feq}, and between
equations \ref{fdef}, and \ref{R-Phi}, except that the indices are paired differently. 
On the other hand, the equations considered bear some similarities with the ones considered for
self-dual Yang-Mills theories with extended supersymmetries \cite{DL92}.

\section{B\"acklund transformations}
\subsection{The principle}
The discussion we are going to present is closely inspired by the corresponding ideas
developed \cite{L91} \cite{DL92}  for self-dual Yang-Mills theories. Nevertheless we are able to
somewhat clarify the mechanism which is at work. At this point let us recall that we want to
establish  transformations between solutions of equations \ref{nonleq}. These coincide with the
symmetric part of equations \ref{dynud} in the gauge where $A_{\overline \mu}=0$ where the first
and third terms on the left are absent.  The new insight is that a  B\"acklund transformation of
the type developed so far \cite{L91} for self-dual Yang-Mills theories,  correspond to a particular
gauge transformation, say
$A\to A'$  (from the on-shell light cone gauge we are using, where
$A_\mu=D_{\overline \mu}\Phi$)  which is such that the symmetric part of equations \ref{dynud}
retain the same form;  so that we may  define a superfield $\Upsilon  $ by letting 
$A'_\mu=D_{\overline \mu}\Upsilon $ thereby obtaining a new solution of equations \ref{nonleq}.    
\subsection{Definition}
Let us thus consider consider a gauge transformation   $A\to A'$ where 
$$
A'_{\alpha}=
SA_\alpha S^{-1}-\left(D_\alpha S\right)S^{-1}
$$
Clearly 
$$
A'_{\mu}= 
 S\left(D_{\overline \mu}\Phi\right)S^{-1}-\left(D_\mu S\right)S^{-1},
\quad 
A'_{\overline \mu}= 
-\left(D_{\overline \mu} S\right)S^{-1}, 
$$
satisfy equations \ref{dynud}. Thus we get, aftere symmetrizing in $\mu$ and $\nu$ 
\beq
D_{ \mu} A'_{\overline \nu}+D_{ \nu} A'_{\overline \mu} 
+\left\{D_{\overline \nu} A'_{ \mu}+D_{\overline \mu} A'_{ \nu}
+\left[A'_{\mu},\, A'_{\overline \nu}\right]_+
+\left[A'_{\nu},\, A'_{\overline \mu}\right]_+
\right\}=-4 A'_8\delta_{\mu,\overline \nu}. 
\label{dynudp}
\eeq
This will have the same form as the corresponding equation for $A_{\overline \mu}$, i.e.
the symmetric part of equation
\ref{dynud0}, if the terms in bracket is also proportional to $\delta_{\mu \nu}$. After some
calculation one sees that this condition is equivalent to the conditions
\beq
 D_{ \mu}  \left(  S^{-1} D_{\overline \nu} S \right) 
+\left[D_{\overline \mu}\Phi,\,  \left(S^{-1}D_{\overline\nu} S\right)\right]_+
+\left\{\mu \leftrightarrow  \nu\right\}
\propto \delta_{\mu,\overline \nu}
\label{Sdef}
\eeq 
If these conditions can be solved, $A'_{ \mu}$ is such that 
$D_{\overline \mu}A'_{ \nu}+D_{\overline \nu}A'_{ \mu}\propto\delta_{\mu, \nu}$. Thus we may
define a superfield $\Upsilon$ by letting $A'_{ \mu}=D_{\overline \mu}\Upsilon$. Thus we finally 
relate $\Phi$ and $\Upsilon$ by the relations 
\beq
S\left(D_{\overline \mu}\Phi \right)S^{-1}+SD_\mu S^{-1}=\partial_{\overline \mu} \Upsilon. 
\label{upsdef}
\eeq
 Since $A'_\mu$ satisfies the
gauge transformed of equation \ref{dynuu}, $\Upsilon$ satisfies the equations 
$$
D_\mu D_{\overline \nu}\Upsilon+D_\nu D_{\overline \mu}\Upsilon + 
\left[D_{\overline \mu}\Upsilon,\,  D_{\overline
\nu}\Upsilon\right]_+=2\delta_{\mu \nu} \left(A'_0+A'_9\right). 
$$ 
We have thus obtained a B\"acklund transformation for equation \ref{nonleq}, since the latter
equations have the same form. 
\subsection{Solving the condition for $S$}
The essential point is that the first order equations \ref{feq} which are equivalent to equations
\ref{nonleq} are precisely such that the conditions \ref{Sdef} may be explicitly solved. 
We shall show this for the particular case of the gauge group $A_1$. 
Let us then write  
$$
\Phi=\Phi_+X^++\Phi_-X^-+\Phi_0H
$$
with $\left[H,\,X^\pm\right]_-=2X^\pm$, $\left[X^+,\,X^- \right]=H$.  Consider the automorphism of
the algebra generated by $r$ which is such that 
$$
rX^\pm r^{-1}=X^\mp,\quad rHr^{-1}=-H. 
$$
First, we change the gauge transformation from $S$ to $Sr$ so that the condition to be satisfied
becomes 
$$
\left[S^{-1}\left(D_{\overline \nu} S\right),\, \left(D_{\overline \mu}\tilde \Phi\right)\right]_+
+\left[S^{-1}\left(D_{\overline \mu} S\right),\, \left(D_{\overline \nu}\tilde \Phi\right)\right]_+
$$
$$
+D_\mu \left(S^{-1}  D_{\overline \nu}S\right)
+D_\nu \left(S^{-1}  D_{\overline \mu}S\right)=0 
$$
where $\tilde \Phi=r\Phi r^{-1}$. A straightforwrd computation shows that these equations are
satisfied by letting 
\beq
S=e^{\tilde \Omega X^+}\left(\Phi_+\right)^H, 
\label{a1res1}
\eeq
if the superfield $\tilde \Omega$ satisfies the equations   
$$
D_{\overline \mu}\tilde \Omega= 
2\Phi^+ D_{\overline \mu}\Phi^0 
+D_\mu \Phi^+.  
$$
The crucial point is that this latter conditions are almost the same as equations \ref{omeq}.
Indeed, it is easy to see that they are satisfied if we let 
\beq
\tilde \Omega=\Omega_++\Phi_0\Phi_+
\label{omtdef}
\eeq
where we have used the decomposition $\Omega=\Omega_+X^++\Omega_0H+\Omega_-X^-$
\section{Outlook} In summary, we have been able to devise a B\"acklund transformation for
equations derived from \ref{dynuu} and the symmetric part of equation \ref{dynud},  in a gauge
where equation \ref{dyndd} becomes trivial. This is yet nother indication that the system of
equations 
\ref{dynuu},  \ref{dyndd},  and the symmetric part of \ref{dynud} is completely integrable and
has similarities with  self-dual Yang-Mills in four dimensions.  Another similar property is
the existence of a Lax representation \cite{G99-1} \cite{G99-2}  analogous to the one of Belavin and
Sakharov.  As a matter of fact, it is easy to see that we have solved a variant of the field
equations,
\ref{dynuu}--\ref{dynud} where the last is replaced by   
$$
D_{ \mu} A_{\overline \nu}+D_{\overline \nu} A_{ \mu} +\left[A_{\mu},\,
A_{\overline \nu}\right]_+=-2\sum_{i=1}^8 A_i\gamma^i_{\mu,\overline \nu}+
\sum_{i,j=1}^7 B_{i j}\gamma^{i j 8}_{\mu,\overline \nu}. 
$$
There appears an additional bosonic superfield $B_{i j}$. The physical meaning of this modified
dynamics is under investigation \cite{GS99}. 

The last point is that one may be worried that the B\"acklund transformation, being based on a gauge
transformation maybe trivial. This is not so because we are able to relate  solutions  which
are both in the same on-shell light cone gauge although our gauge transformation does not respect
this gauge condition. This is possible since, after transformation, the term in
bracket in equation \ref{dynudp} may be  lumped into the right hand side. Thus in effect, the
solutions are not related by gauge transformations. 

\noindent {\bf Acknowledgements.} I am indebted to D. Fairlie for discussions and for pointing
earlier references \cite{L91}  \cite{DL92}.  
 

\fin